\documentclass[osajnl,a4paper,twocolumn]{revtex4}  

\usepackage{graphicx}

\begin{document}

\title{Numerical heterodyne holography with two-dimensional
photodetector arrays}

\author{F. Le Clerc, L. Collot}

\address{
Thomson CSF Optronique, Rue Guynemer B.P. 55, 78 283 Guyancourt, France\\
}
\author{M. Gross}

\address{
Laboratoire Kastler-Brossel, UMR 8552 CNRS, Ecole Normale Sup\'{e}rieure, Universit\'{e} Pierre et Marie Curie, 24 rue Lhomond 75231 Paris cedex 05 France\\
}

\begin{abstract}
We present an original heterodyne holography method for digital holography that relies on two-dimensional
heterodyne detection to record the phase and the amplitude of a field. The technique has been tested on objects as much as 13 mm in size. Consistency checks were performed, and high-resolution images were computed.
We show the requirement for a spatial filter to select properly sampled near-axis photons. Heterodyne
holography is superior to off-axis digital holography for both field of view and resolution.
\end{abstract}


\maketitle

As was demonstrated by Gabor \cite{gabor1949microscopy}, in the early 1950.s,
the purpose of holography is to record the phase and
the amplitude of the light coming from an object under
coherent illumination. Classical holography does
not provide straightforward access to the holographic
data. For quantitative analyses of those data, in digital
holography \cite{Kreis1998principles} (DH) photographic films were replaced
by two-dimensional electronic detectors. In both digital
and thin-film holograms a ghost field and the remaining
part of the reference field are superimposed
upon the reconstructed object field \cite{cuche1998numerical}. A solution to this
problem is to tilt the reference beam with respect to the
object \cite{Leith1962reconstructed}, to separate physically the spatial-frequency
components. Although this off-axis technique is acceptable
for high-resolution holographic films, it is
hardly compatible with the limited resolution of digital
holograms. All these artifacts arise from the fact
that among all these methods only one field quadrature
is measured. Measuring both quadratures requires
recording at least two interferograms with distinct reference
phases \cite{schnars1994direct}. In classical holography this is done
with a thick plate that records several intensity fringes
in depth.

In this Letter we describe a heterodyne holographic
scheme in which the reference beam is dynamically
phase shifted with respect to the signal field. This
shift produces time-varying interferograms on a two-dimensional
sensor. In our experiment the phase shift
is linear in time (frequency shift). Intensity $I$ in the
detector plane results from the interference of the
signal field with the $\delta f$-shifted reference field:
\begin{equation}\label{Eq_1}
   I(t)=\left|E_S+E_R \exp(2i\pi \delta f t)\right|^2
\end{equation}
where $E_S$ and $E_R$ represent the complex amplitudes
of the signal and the reference fields, respectively.
$L$ intensity $I_l$	 ($l=0...L-1$)	  measurements are
performed within a $\delta f$ period at $t_l=2\pi l / \delta f$. We
obtain $E_S$ by demodulating $I$ :
\begin{equation}\label{Eq_2}
   E_S=\left( \frac{1}{L E_R^*}\right)\sum_{l=0}^{L-1}I_l \exp i\left(\frac{2\pi l}{L} \right)
\end{equation}
where $^*$ is the complex conjugate. For $L=4$, $E_S$ is
proportional to $(I_0-I_2)+i(I_1-I_3)$. Heterodyne
holography (HH) thus measures the phase, using the
information obtained at different times, and DH extracts
the phase from measurements made of different
pixels \cite{schnars1996digital}. In both cases the sampling theorem  \cite{Kreis1998principles} restricts
the largest admissible field-of-view angle $\theta$ to
\begin{equation}\label{Eq_3}
  |\theta| \leq  \theta_{max} =\lambda/(2 d_{pixel})
\end{equation}
where $\lambda$ is the wavelength and  $d_{pixel}$
 is the pixel spacing.
In DH the object must be off axis, which yields
the constraint [3] that $\theta \geq \theta_{min}= N d_{pixel}/(2 D)$, where $N$
is the number of pixels and $D$ is the sensor-to-object
distance. The DH field of view, $\theta_{min} \leq \theta \theta_{max} $, is
thus much smaller than for HH: $|\theta| \leq  \theta_{max}$. Whereas
$\theta_{min} \leq  \theta_{max}$.	
 restricts DH to the far field with respect
to $D_{min}=N d_{pixel}/\lambda $, HH works in both the near field and the far field modes. In the far field, both
HH and DH angular resolution reaches the diffraction
limit, $\theta_{diff}=\lambda/(N d_{pixel})$, which corresponds to the two dimensional
sensor's size. In HH the number of resolved
pixels (i.e., field of view/angular resolution) is
N. This remains true in the near-field regime.

\begin{figure}
\begin{center}
  \includegraphics[width=8 cm]{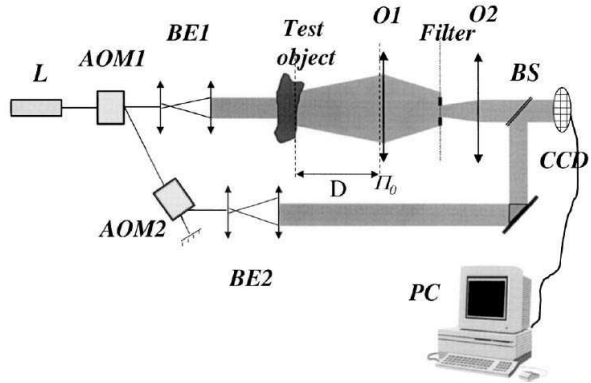}
  \caption{Experimental setup: L, He.Ne laser; other abbreviations
defined in text.}\label{Fig_1}
\end{center}
\end{figure}

Although HH works on ref lection, we focus here
on a transmission configuration. The general setup,
shown in Fig. \ref{Fig_1}, is a Mach.Zehnder interferometer
composed of two beams (reference and signal) from
the same coherent source (a 5-mW He.Ne laser). The
two beams are expanded by beam expanders BE1 and
BE2 and combined by a beam splitter (BS) on a CCD
camera with $N_x=768$ and $N_y= 576$   pixels,
$d_{pixel_x}=8.6 ~\mu$m	and $d_{pixel_y}=8.3 ~\mu$m.
 All $x$ and $y$ subscripts
below refer to $x$
 or $y$ axes, respectively. We $\delta f$ shift
the reference beam by combining two acousto-optic
modulators, AOM1 and AOM2, working at
$\Delta f + \delta f$ and $-\Delta f$, respectively, with
$\Delta f = 80$ MHz. $ \delta f = 6.25$ Hz
is equal to one quarter of the CCD image frequency
$f_I=25$  Hz ($L=4$). The video frame is acquired by
an 8-bit analog frame grabber (Matrox Meteor), and a
Pentium II 450-MHz computer calculates the complex
field in real time.

\begin{figure}
\begin{center}
  \includegraphics[width=8.5 cm]{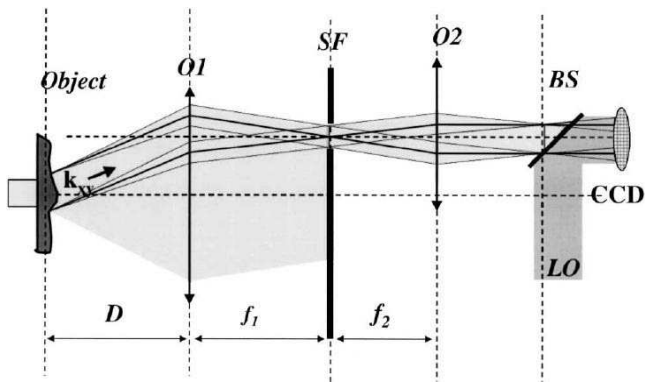}
  \caption{SF system: LO, local oscillator; other abbreviations
defined in text.}\label{Fig_2}
\end{center}
\end{figure}

To select the signal on-axis photons, we insert between
the object and the BS a spatial filter system
(Fig. 2) composed of two confocal objectives, O1 (focal
length, $f_1=50$ mm) and O2 ($f_2=25$ mm), with
a spatial filter (SF) in their common focal plane.
O1 transforms the field into its $k$-space components
in its focal plane where the SF selects the photons
that fulfill Eq.\ref{Eq_3}. O2 backtransforms the field to
real space. The SF, O2, the BS, and the CCD are
kept in alignment, so the selected photons reach the
CCD nearly parallel to the reference beam. Equation
\ref{Eq_3} yields a rectangular SF of dimensions
$d_{x,y}=2 \tan (\lambda/2 d_{pixel_{x,y}}) f_2$
($d_x=1.84$ mm and $d_y=1.90$ mm).
The O1-O2 optical system magnifies the incoming
beam by a factor $f_1/f_2=1/2$. Our setup thus has
$x$ and $y$ fields of view of $\pm 1.05^\circ$ and $\pm 1.09^\circ$ , corresponding
to an equivalent CCD with magnified pixels
$d'_{pixel}= 2 d_{pixel}$.
O1 is fixed, and the SF, O2, the
BS, and the CCD camera can be $x$
 and $y$ translated by
step motors. When the SF is not centered on the optical
axis of O1, the system records holograms that correspond
to tilted $k$ components, as depicted in Fig. 2.
Those displacements allow us to center the SF accurately.
This feature will be used in the future to record
wider-field-of-view holograms by merging several holograms
with different $k$-component tilts.

\begin{figure}
\begin{center}
  \includegraphics[width=8.5 cm]{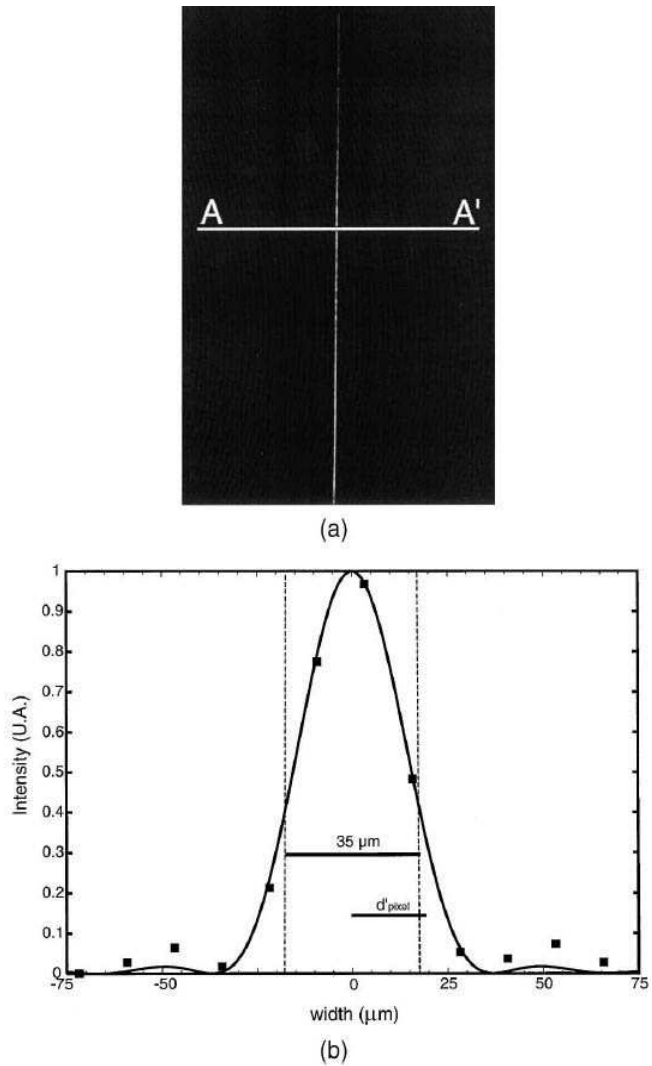}
  \caption{(a) Contact image of the slit (b) cut along
$A A'$: filled squares, experimental intensity points; dashed
lines, initial slit; solid curve, theoretical intensity profile.}\label{Fig_3}
\end{center}
\end{figure}

To test our setup quantitatively, we recorded the
hologram of a narrow slit 9.5 mm high and $w=35 \mu$m
wide at $D=60$ cm with a 480-ms integration time.
We got $E_S(x,y,z=0)$ on the equivalent CCD. To
compute $E_S(x,y,z=D)$  on the slit we considered the
Fresnel propagation of $E_S$ from $z=0$ to $z=D$, which
can be formally expressed as an $x$
 and $y$ convolution
product (symbol $\otimes$):
\begin{eqnarray}\label{Eq_4}
\nonumber  E_S(x,y,D) &=& e^{i k_0 D}\left[P(z=D) \otimes  E_S(x,y,0)\right]\\
  P(x,y,z) &=& \frac{1}{i\lambda z} \exp\left[ i\frac{k_0}{2z}\left(x^2+y^2\right)\right]
\end{eqnarray}
We obtained a 13.2 mm $\times $ 9.6 mm (equivalent CCD
size) contact image of the slit in its plane [Fig. \ref{Fig_3}(a)].
The $z=0$ field is calculated on a $1024 \times 1024$ grid
by bilinear interpolation of the $768 \times 576$ experimental
points, and the convolution product is calculated by the
fast-Fourier-transformmethod \cite{schnars1994direct}, with the grid size kept
constant over z. Figure \ref{Fig_3}(b) shows the intensity along
the $A A'$ cut. Points are experimental data on the
fast-Fourier-transform grid. The solid curve is the
calculated diffracted intensity of an ideal wide slit
(dashed lines). Both diffraction and spatial averaging
over the magnified pixels are calculated; the agreement
is good. $D$ is larger than $D_{min_x} = 35.9$ cm and
$D_{min_y} = 25.0$ cm, so diffraction is the main contributor
to resolution; here it is assumed to be equal to the distance
that yields a sharp contact imaging. Comparing
$D$ with the O1-to-object distance, we find that the
equivalent CCD is located $\sim 27$ cm behind O1. This
location does not depend on $D$ and agrees with the positions
of O1, O2, and the CCD.

\begin{figure}
\begin{center}
  \includegraphics[width=8 cm]{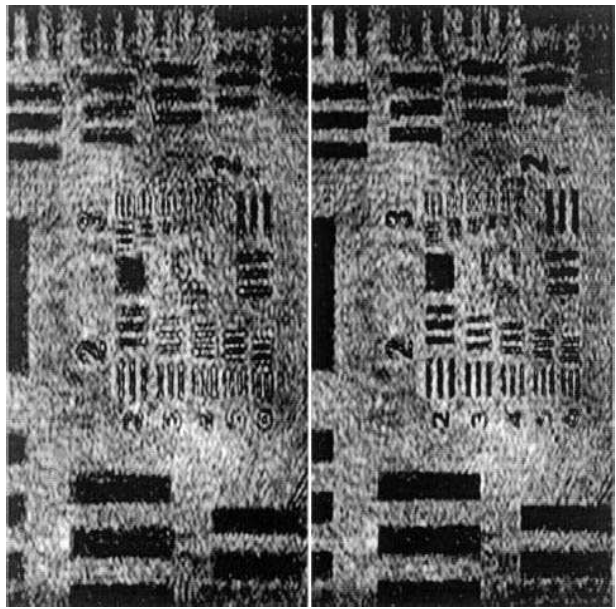}
  \caption{Contact image of a U.S. Air Force test chart target
at $D=58$ cm. Right, image on focus; left, 1.2-mm out-of-focus
image.}\label{Fig_4}
\end{center}
\end{figure}

Figure \ref{Fig_4} shows the contact image of a transmission
U.S. Air Force test chart target at $D=58$ cm. At the
right, Fig. \ref{Fig_4} is exactly on focus, whereas at the left the
target is 1.2 mm out of focus. The blur is due to a
defocus equal to the $x$ depth of focus
(DOF): $DOF_{x,y}=\lambda(1-NA_{x,y}^2)^2/NA_{x,y}^2)^2$ ($DOF_x=1.22$ mm, $DOF_x=2.32$ mm).
Here $NA_{x,y}=N_{x,y} d'_{pixel_{x,y}}/D$ is the numerical aperture.

\begin{figure}
\begin{center}
  \includegraphics[width=8 cm]{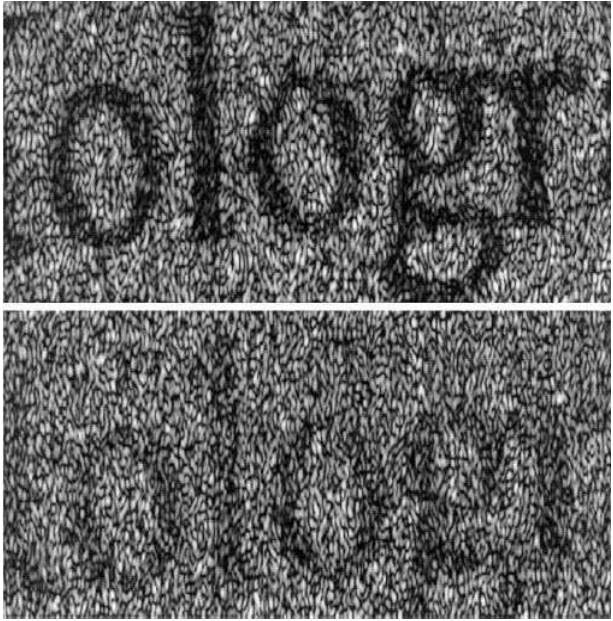}
  \caption{Contact images of black characters printed on
diffusive plastic sheets with (top) and without (bottom)
a SF.}\label{Fig_5}
\end{center}
\end{figure}

Figure \ref{Fig_5} shows the central part
of the contact images of black characters printed on
a diffusive plastic sheet at $D=30$ cm with (top) and
without (bottom) a SF. Here the plastic substrate
scatters light to angles larger than $\theta_{max}$	. Without a
SF the off-axis photons scramble the useful information,
reducing the image contrast. This phenomenon
is a spatial-frequency-folding effect.

As was shown in this Letter, HH measures the field
within the SF sampling cone without information loss.
HH has many other advantages. By measuring the
field in a $z=0$ plane we are able to compute it at
any $z$ plane, thus getting three-dimensional information.
Because the detected interference arises from
the heterodyne mixing of a strong reference with a
weak signal, HH is expected to be sensitive up to the
photon level. Compared with intensity imaging, HH
offers a huge dynamic range because it detects amplitude.
The time-varying character of HH provides good
immunity against any dc and nonharmonic (with respect
to $\delta f$) errors. The CCD image phase shift ($90^\circ$)
is highly accurate because it results from a frequency-offset
$\delta f$. It is also possible to shift the heterodyne frequency
$\delta f$ for sideband heterodyne detection. Making
$\delta f= (1/4) f_I + f_{mod}$ allows the vibration of an object
at $f_{mod}$ to be detected. HH works with a low-power
cw laser. HH may be performed with low coherent
laser sources to select a narrow space slice on the object
where coherence is conserved.

Within the numerous possible applications of the
method, we have explored diffusing media \cite{gross1999imagery} and aperture
synthesis applications, and further experiments
are in progress. In both cases, a SF is useful.

We thank Thomson-CSF Optronique for its support
and C. Boccara for fruitful discussions. This
research was supported by the French Direction Generale
de l'Armement under contract 98 10 11A.000.

F. Le Clerc's e-mail address is leclerc@lkb.ens.fr.


\begin{thebibliography}{0}
\expandafter\ifx\csname natexlab\endcsname\relax\def\natexlab#1{#1}\fi
\expandafter\ifx\csname bibnamefont\endcsname\relax
  \def\bibnamefont#1{#1}\fi
\expandafter\ifx\csname bibfnamefont\endcsname\relax
  \def\bibfnamefont#1{#1}\fi
\expandafter\ifx\csname citenamefont\endcsname\relax
  \def\citenamefont#1{#1}\fi
\expandafter\ifx\csname url\endcsname\relax
  \def\url#1{\texttt{#1}}\fi
\expandafter\ifx\csname urlprefix\endcsname\relax\def\urlprefix{URL }\fi
\providecommand{\bibinfo}[2]{#2}
\providecommand{\eprint}[2][]{\url{#2}}

\end{thebibliography}


\begin{thebibliography}{1}

\bibitem{gabor1949microscopy}
D.~Gabor.
\newblock Microscopy by reconstructed wave-fronts.
\newblock {\em Proceedings of the Royal Society of London. Series A,
  Mathematical and Physical Sciences}, pages 454--487, 1949.

\bibitem{Kreis1998principles}
T.M. Kreis, W.P.O. J{\"u}ptner, and J.~Geldmacher.
\newblock Principles of digital holographic interferometry.
\newblock In {\em Proceedings of SPIE}, volume 3478, page~45, 1998.

\bibitem{cuche1998numerical}
E.~Cuche, P.~Poscio, and C.D. Depeursinge.
\newblock Numerical holography with digital recording devices.
\newblock In {\em Proceedings of SPIE}, volume 3196, page~24, 1998.

\bibitem{Leith1962reconstructed}
E.N. Leith and J.~Upatnnieks.
\newblock Reconstructed wavefronts and communication theory.
\newblock {\em J. Opt. Soc. Am.}, 52:1123--1128, 1962.

\bibitem{schnars1994direct}
U.~Schnars.
\newblock Direct phase determination in hologram interferometry with use of
  digitally recorded holograms.
\newblock {\em JOSA A}, 11(7):2011--2015, 1994.

\bibitem{schnars1996digital}
U.~Schnars, T.M. Kreis, and W.P.O. J{\"u}ptner.
\newblock Digital recording and numerical reconstruction of holograms:
  reduction of the spatial frequency spectrum.
\newblock {\em Optical Engineering}, 35:977, 1996.

\bibitem{gross1999imagery}
M.~Gross, F.~Le Clerc, and L.~Collo.
\newblock {\em Waves and Imaging through complex media}, chapter Imagery of
  Diffusing Media by Heterodyne Holography.
\newblock Kluwer Academic: P. Sebbah Ed., 2001.

\end{thebibliography}

\end{document}